\documentclass[preprint]{revtex4-2}

\usepackage{graphics}
\usepackage{subcaption}
\usepackage{graphicx}
\usepackage{dcolumn}
\usepackage{bm}

\begin{document}

\title{Compact spectrometer based on disordered multi-mode interferometer}
\author{Ankit Poudel}
\email{apoudel@pinstitute.org}
\affiliation{Biophotonics Lab, Phutung Research Institute, PO Box 12335, Tarakeshwor-5, Kathmandu, 44611, Nepal}
\author{Pravin Bhattarai}
\affiliation{Biophotonics Lab, Phutung Research Institute, PO Box 12335, Tarakeshwor-5, Kathmandu, 44611, Nepal}
\author{Rijan Maharjan}
\affiliation{Biophotonics Lab, Phutung Research Institute, PO Box 12335, Tarakeshwor-5, Kathmandu, 44611, Nepal}
\author{Richard J Curry}
\affiliation{Photon Science Institute, Department of Electrical and Electronic Engineering, University of Manchester, M13 9PL, UK}
\author{Iain F Crowe}
\affiliation{Photon Science Institute, Department of Electrical and Electronic Engineering, University of Manchester, M13 9PL, UK}
\author{Ashim Dhakal}
\email{ashim.dhakal@pinstitute.org}
\affiliation{Biophotonics Lab, Phutung Research Institute, PO Box 12335, Tarakeshwor-5, Kathmandu, 44611, Nepal}

\date{\today}

\begin{abstract}
We demonstrate a compact (40 ${\mu}$m $\times$ 260 ${\mu}$m) spectrometer based on multimode interference aided by scattering of light from random SiO$_2$-filled hole arrays on a silicon-on-insulator platform. We characterize the performance of the spectrometer for wavelengths around 1310 nm, and report that the spectrometer can reconstruct a broadband $\sim$ 67 nm source, as well as Lorentzian probes of $\sim$ 1 nm bandwidth. This compact nanometer level resolution spectrometer can be fabricated at a low cost for lab-on-a-chip sensing and imaging applications.
\end{abstract}
\maketitle

\section{Introduction}
Spectrometers are a widely used, indispensable tool for the measurement of light scattering and emission across a broad range of disciplines, in material science, chemical and biological sciences, and astronomy. However, current state-of-the-art high-resolution spectrometry relies on bulky and expensive bench-top systems, making them inaccessible to many application settings. Recent developments in mass-scalable, silicon photonic integrated circuits (PICs) have enabled the realization of compact and lower-cost 'lab-on-a-chip' devices, operating over a wide optical wavelength range. By leveraging the small footprint of these PICs, various high-performance spectrometers integrated on the silicon platform~\cite{cheben2007high, ryckeboer2013silicon, ma2018high, xia2011high} have now been demonstrated. The traditional approach has been to disperse the incident light into an array of filters or gratings and record the spectral contents using an array of photo-detectors. Since the resolution scales with the linear dimension of these types of grating-based spectrometers, the device footprint tends to be relatively large. Recent examples are an echelle grating-based spectrometer with a footprint of 3 mm $\times$ 3 mm and an arrayed waveguide grating (AWG)-type spectrometer with a footprint varying from $\sim$1 cm$^2$ to $\sim$1 mm$^2$  depending on refractive index contrast of the PIC platform~\cite{ma2018high, janz2004planar, zirngibl1992demonstration, he1998monolithic}. A number of devices with improved resolution and/or smaller footprint, based on ring resonator arrays and photonic crystal defect cavities have also been demonstrated, but these proved to be highly sensitive to fabrication errors~\cite{xia2011high}.

In addition to these more standard device architectures, the incorporation of disordered structures in the  spectrometer design has also been explored~\cite{redding2013compact, hartmann2020waveguide}. Redding \textit{et al.} presented such a device in which a semi-circular structure with a disordered array of air-filled holes acts as a diffuse component, surrounded by a full-bandgap photonic-crystal boundary to channel the light onto the detector. This device was able to achieve 0.75 nm resolution at a wavelength of 1500 nm with a 25 nm bandwidth~\cite{redding2013compact}. Extending the operational wavelength range, a similar PIC-based device was demonstrated by Hartmann \textit{et al.} on a silicon nitride (Si$_3$N$_4$) platform, with successful reconstruction of several signal probes\cite{hartmann2020waveguide}. Although these approaches can be useful in various spectroscopy applications, they have tended to require a complexity of components (such as photonic-crystal waveguide channels) and high-resolution fabrication (such as electron beam lithography, EBL), both of which increases the fabrication cost and complexity precluding mass-scale fabrication.

In a channel-waveguide-based photonics framework, speckle patterns due to multimode interference can be used to reduce the size and complexity, and enhance the bandwidth of operation.  We reported the design of such disordered spectrometer by incorporating randomized holes within the multi-mode interferrometer (MMI)\cite{bhattarai2020mmi}. These devices permit higher fabrication tolerance, therefore allows for a lower-cost UV lithography suitable for mass-fabrication of the devices. Our design is based on analysis of the speckle patterns formed as transmitted light passing through the MMI with scattering holes providing a unique fingerprint of spectral pattern of the wavelengths of an input probe signal. These wavelength-dependent speckle patterns are measured and stored in a transmission matrix describing the spectral-to-spatial mapping of the spectrometer which can be used to reproduce any arbitrary probe signal. The speckle signal intensity from the disordered region of the device can be expressed as:
\begin{equation}
    I(r)=\int T(r, \lambda) S(\lambda) d\lambda
\end{equation}

where $T(r, \lambda)$ is transmission function dependent on the position, $r$, and wavelength, $\lambda$, and $S(\lambda)$ is the spectral flux density of the input signal being probed. In matrix algebra formalism, a discretized transmission matrix, $T$ can be obtained by discretization of the spectral and spatial components, which yields:

\begin{equation}
    [I]_{M\times N}=[T]_{M\times N}.[S]_{N\times N}
\end{equation}

where $I$ is a vector representing the intensities measured in $M$ spatial channels of the output, and $S$ the intensities in $N$ spectral channels from the input\cite{redding2013all}. 

Following on from our original simulation work, here we experimentally demonstrate the operation of a compact spectrometer device built around this design. This spectrometer is based on a simple $1\times16$ MMI splitter with randomized holes on the silicon-on-insulator (SOI) platform and operates in the near-infrared regime $\sim$ 1310 nm, key to several
applications; methane sensing \cite{cubillas2008methane}, optical coherence tomography (OCT) based imaging\cite{maharjan2021non} and optical gyroscopes\cite{dell2014recent}. We first calibrated the transmission matrix $T$ over the full wavelength range of our swept source (1281 nm to 1348 nm) and used this to create a generic reconstruction algorithm based on a truncated inversion technique\cite{redding2013all}, capable of accurately reconstructing the spectra from any signal source with an operating wavelength in this band.

\section{Devices and methods}
\subsection{Device design and fabrication}

\begin{figure}[ht]
\centering\includegraphics[width=0.8\textwidth]{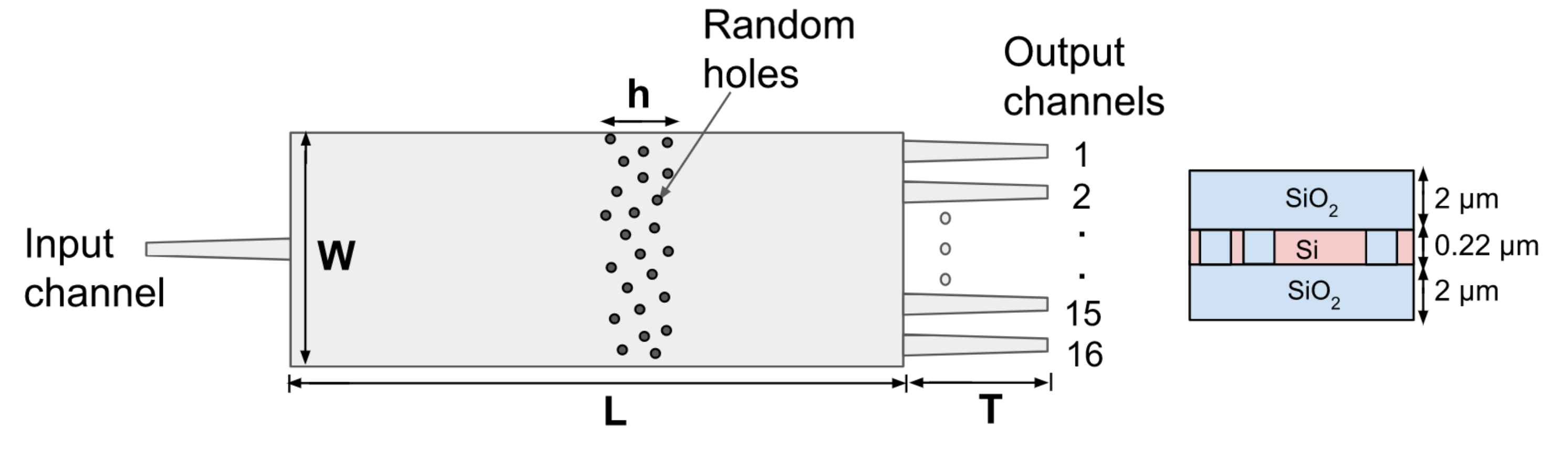}
\caption{Schematic representation of $1\times16$ random scattering MMI based spectrometer where, $L = 260~\mu m$, $W = 40~\mu m$, $T = 20 ~\mu m$, $h = 15 ~\mu m$.}
\label{fig:Illustration}
\end{figure}

We designed the spectrometer around the commercial standard silicon-on-insulator (SOI) starting wafer, with a $220 \pm 20$ nm silicon layer atop a 2 $\mu$m SiO$_2$ buried oxide (BOX) on a silicon substrate. The devices were fabricated via UV lithography at Southampton University under the CORNERSTONE program. A post-fabrication protective layer of SiO$_2$ was deposited above the device layer. Figure~\ref{fig:Illustration} illustrates the basic design of our disordered MMI spectrometer device, consisting of a single channel on the input side into which light is launched and 16 output channels from where the diffuse light is collected. Vertical grating couplers were included in the device design, based on an optimised coupling angle of between 9$^{\circ}$ and 11$^{\circ}$(off-normal) for efficient coupling of light into the guided mode. Such grating couplers are now a fairly standard feature in PICs owing to the straightforward (alignment tolerant) fiber-to-chip delivery and collection of light for rapid, wafer scale testing. In order to generate the randomized hole pattern in our structure, we used the Bridson algorithm to achieve a uniform blue noise distribution~\cite{bridson2007fast}. This two-dimensional scattering structure consists of an array of randomly positioned SiO$_2$ filled holes, each of 0.4 ${\mu}$m diameter, that acts as a diffusive regime in the silicon slab of the MMI body. We used a 2D-finite difference time domain (FDTD) solver \textit{(ANSYS/Lumerical solutions)} to simulate the following device design parameters; overall MMI device geometry and the number, position, pitch and width of the spatially randomized holes. These parameters were optimized, to enable us to achieve a similar operational-bandwidth ($\sim$ 67 nm) to that of our swept laser source.

\subsection{Experimental Setup}
\label{sec:experimental setup}

\begin{figure}[ht]
\centering\includegraphics[width=0.8\textwidth]{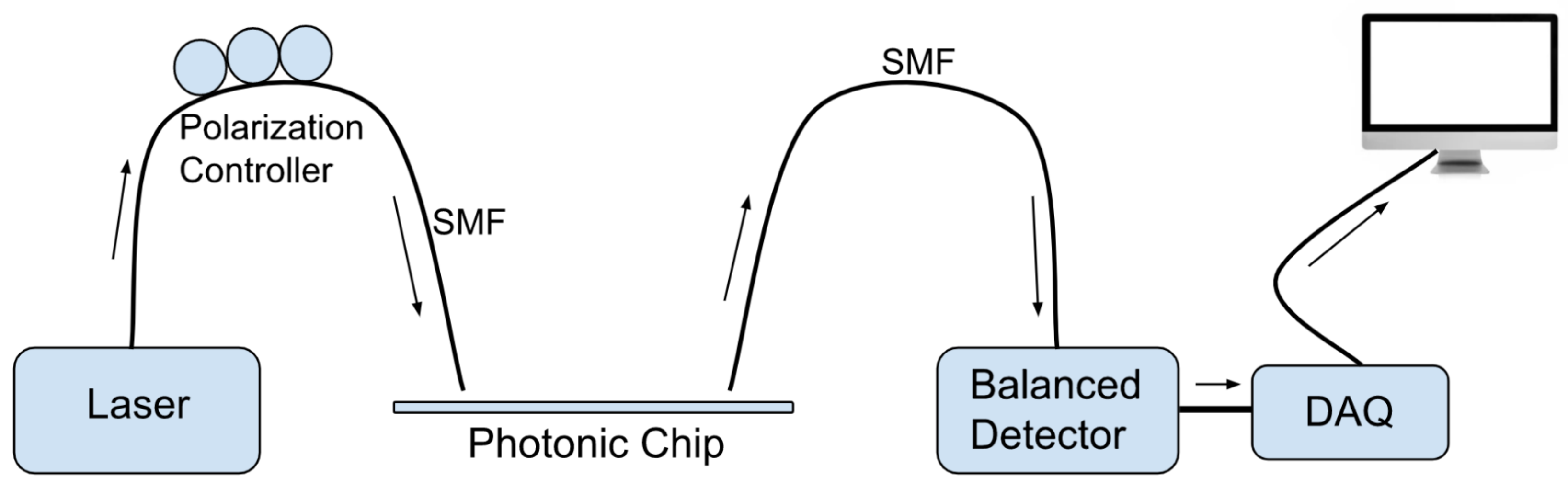}
\caption{Schematic of the experimental setup.}
\label{fig:Schematic}
\end{figure}

Figure~\ref{fig:Schematic} shows the experimental setup for the characterization of our spectrometer device. For initial calibration of the device, we used a 100 nm bandwidth swept laser source \textit{(Santec HSL-20)} with a center wavelength of 1310 nm and average optical power of 20 mW, and 100 kHz sweeping rate. The laser can act as a broadband light source, when a slow (<1 ms) detectors are used or when the power is averaged over < 1 ms time. Single mode fibers were used to couple the light into and out of the PIC. Precise fiber-to-chip alignment was achieved by mounting the cleaved ends of the fibers to $xyz$ translation stages \textit{(Thorlabs NanoMax313D)}. We used an \textit{InGaAs} detector \textit{(Thorlabs S154C)} mounted on a power meter \textit{(Thorlabs PM100D)} to calibrate the optical alignment by maximizing the power collected at each detector before taking the spectral measurements. Spectral data was acquired using a balanced photo-detector \textit{(Thorlabs PDB430C)} and a high speed data acquisition (DAQ) card \textit{(AlazarTech ATS 9350)} at 500 MS/s for each output channel.

\subsection{Characterization methodology}
\begin{figure}[ht]
\centering\includegraphics[width=4in]{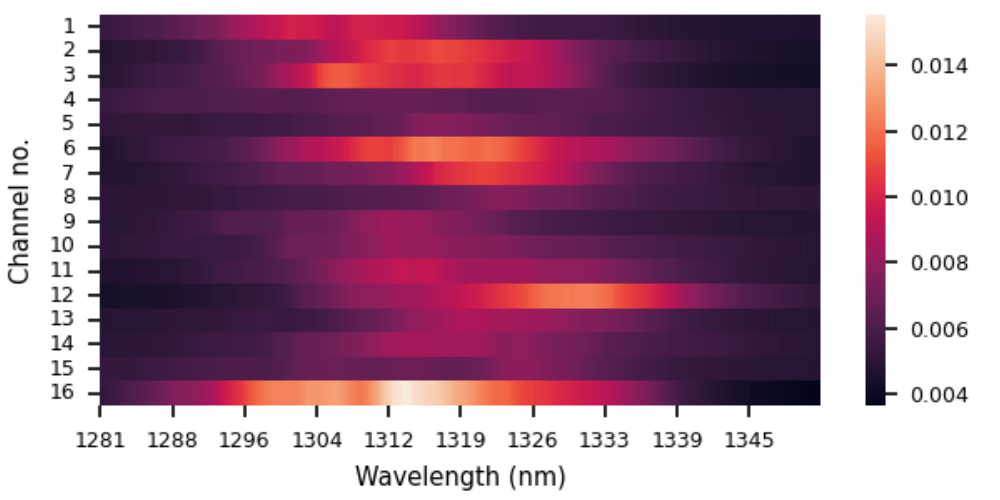}
\caption{The measured intensity distribution on the detection channels as a function of the input wavelength.}
\label{fig:Speckle}
\end{figure}

For the spectral characterization of the spectrometer, we first used the aforementioned swept laser source and recorded the wavelength-dependent intensity distribution for each of the 16 output channels. Figure~\ref{fig:Speckle} shows the speckle patterns recorded from the device where it is evident that a small shifts in the input wavelength has a notable effect on the transmitted intensities. Once this intensity distribution was recorded, we discretized the source spectral flux density across the wavelength range, 1281 nm to 1348 nm by dividing this into 1100 spectral channels. Our choice of 1100 discrete spectral channels was derived from minimum spectrum reconstruction error and a grating coupler limited bandwidth of  $\sim$ 67 nm around the centre wavelength, which leads to a spectral discretization of $\sim$ 0.06 nm. After discretizing the spectral channels of the input $S$ we stored the calibration in the transmission matrix, $T = IS^{-1}$ where, $I$ is a vector representing the wavelength-dependent intensities measured in 16 spatial channels of the output \cite{redding2013all}. Each column in $T$-matrix represents the intensities measured at the 16 spatial channels for a given spectral channel. 

After calibrating the transmission matrix, we can reconstruct any arbitrary probe spectrum by measuring the wavelength-dependent intensity distribution matrix $I$ on each of the detectors and multiplying by the inverse of the previously calibrated transmission matrix as $S=T^{-1}I$. To obtain $T^{-1}$, we initially used the singular value decomposition technique \cite{redding2013all}, although in practice, we found this type of matrix inversion process to be overly error prone due to experimental noise. Thus, to improve the accuracy of the spectral reconstruction algorithm, we used the truncated inversion matrix method\cite{redding2013all}. 

In order to verify the spectrometer performance, we used one device to generate the $T$-matrix, and another copy device in another chip to perform measurements at the output for source reconstruction. To compare the reconstructed spectra with that of true source, we measured the spectrum using a reference waveguide (4 mm) terminated with identical grating couplers.

In addition to the reconstruction of the experimental broadband spectrum, we also numerically generated single and multiple Lorentzian probe signals of varying spectral bandwidths and varying peak separations, and calculated the speckle patterns using the $T$-matrix in order to test reconstruction capabilities and calculate the spectrometer resolution limit.

\section{Results}

\begin{figure}[ht]
\centering\includegraphics[width=4in]{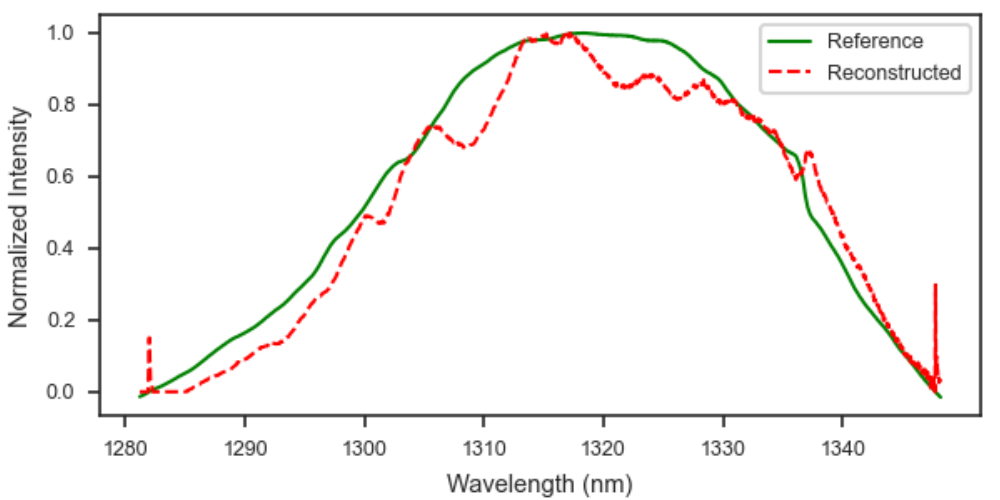}
\caption{Reconstructed spectrum for a continuous, broadband probe signal.}
\label{fig:source reconstruction}
\end{figure}

Figure~\ref{fig:source reconstruction} shows the reconstruction of the continuous broadband spectrum in comparison with the measured reference waveguide spectrum in normalized units (necessary to account for the different losses associated with the reference waveguide and the 16-port MMI). There is very good agreement between the reference and the reconstructed spectra, overall the full 67 nm bandwidth with a normalized Root Mean Square Error (RMSE) of $\sim$ 0.079. The reconstruction error are most likely due to fabrication variations between chips and grating-couplers within the same chip.

We also tested the algorithm with numerically generated Lorentzian signals - which simulate the spectral shape of our swept source laser source, using the $T$-matrix generated from the physical device. We used Lorentzian probes of various full-width at half-maxima (FWHM) in order to determine the bandwidth limit of the device and varied the peak separation of the Lorentzian doublets to determine the resolution limit using Rayleigh-like criteria (the maxima of one peak coinciding with the minima of a nearby peak).

\begin{figure}[ht]
\includegraphics[width=1\textwidth]{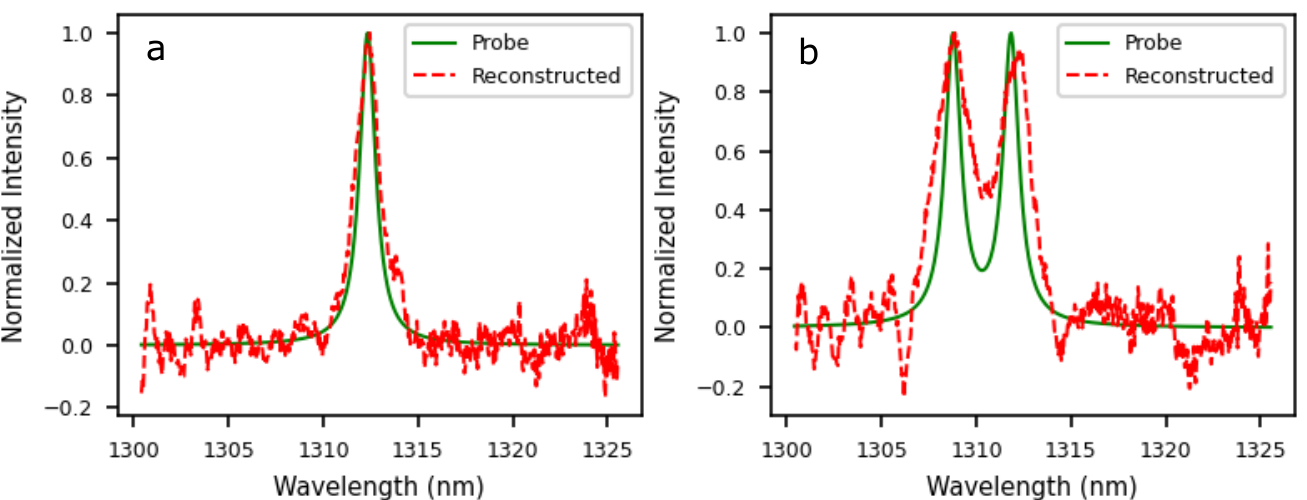}
\caption{(a) Reconstructed Lorentzian peak probe of FWHM at 1 nm. (b) Reconstructed Lorentzian peak probes separated by $\Delta\lambda$ = 3 nm.}
\label{fig:simulated single double}
\end{figure}

Figure~\ref{fig:simulated single double}a shows the spectrum at the bandwidth limit with a FWHM $\approx$ 1 nm, for a single Lorentzian peak centered at a wavelength 1312 nm, with an RMSE of 0.069. Figure~\ref{fig:simulated single double}b shows the reconstruction of a resolution limited Lorentzian doublet signal with FWHM of $\sim$ 1 nm (for each of the underlying Lorentzian signals). In this case, with a peak separation, $\Delta\lambda$ = 3 nm, the RMSE is 0.152.  

\begin{figure}[ht]
\includegraphics[width=1\textwidth]{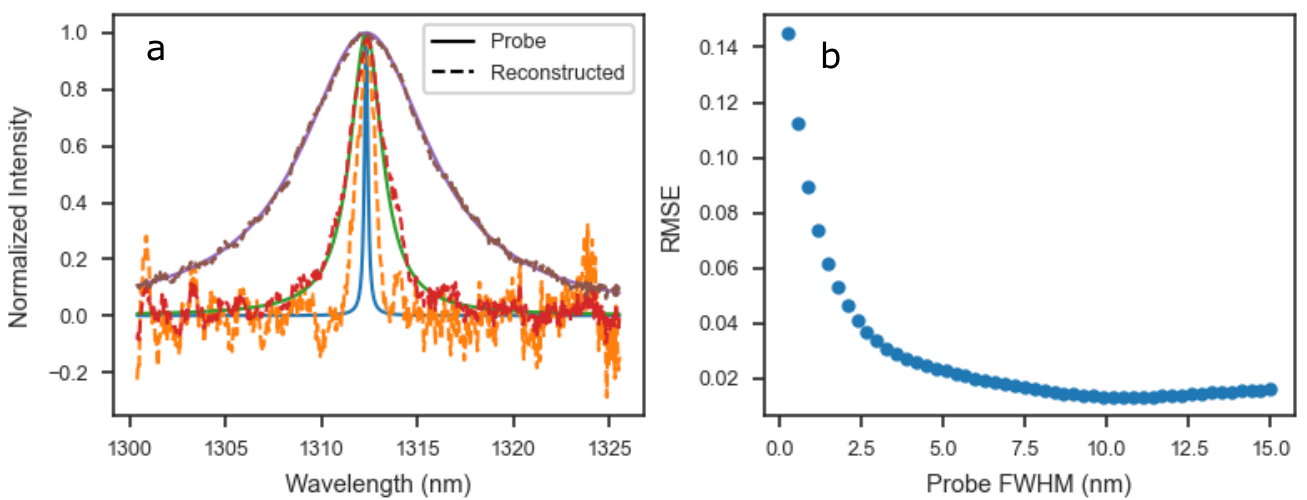}
\caption{(a) Reconstructed spectra (dashed-lines) of varying signal FWHM at 0.25, 2.5, and 10 nm. (b) Reconstruction error (RMSE) vs FWHM (nm).}
\label{fig:RMSE vs width}
\end{figure}

We note that while the bandwidth limit for the single Lorentzian signal was found to be $\sim$ 1 nm, the RMSE increases significantly with the inverse bandwidth. This is illustrated, for three different Lorentzian signals in Fig~\ref{fig:RMSE vs width}a, with FWHM at 0.25 nm, 2.5 nm and 10 nm, all centered on 1312 nm. Figure~\ref{fig:RMSE vs width}b reveals the full extent of the RMSE as a function of input signal bandwidth.

\section{Discussion and conclusions}
In this work we have shown that a disordered $1\times16$ MMI based photonic structure can be used as a compact chip-based spectrometer. The MMI effect combined with enhanced optical path length (OPL) created by an array of spatially randomized scattering SiO$_2$-filled holes in the transmission path of the device, enables nanometer scale resolution in an extremely small device footprint. With this device, we were able to demonstrate spectral reconstruction of a broadband source with RMSE of $\sim$ 0.079 by calibrated $T$-matrix and using a truncated pseudo-inversion algorithm.

The generic reconstruction algorithm was applied to both broadband signal as well as narrowband signals, achieving a bandwidth of 67 nm, a lower bandwidth limit of $\sim$ 1 nm and peak-to-peak resolution of 3 nm. Finally, we examined devices with an increased number of scattering centers (holes), which should theoretically improve resolution (because this scales with OPL). However, we found that this significantly increased optical loss in the device for the feature-size limit of 300 nm for the Multi-Project-Wafer batch of this device, meaning that, in our experimental arrangement, the in-plane optical signal at the output channels are weak for proper spectral analysis. Nevertheless, the generic reconstruction algorithm proves to be robust over the experimental range of the input signals we tested, even in the presence of experimental noise, and we anticipate that this type of device can add new functionality to the silicon photonics 'tool-kit' for a range of nanophotonic, PIC-based 'on-chip' spectroscopy applications.

\section*{Acknowledgements}
This work is partially supported by EPSRC grants EP/R014418/1 and EP/V001914/1, and TWAS/SIDA/UNESCO grants: 18-013 RG/PHYS/AS\_I, 21-334 RG/PHYS/AS\_G and 20-278 RG/PHYS/AS\_G. We thank ANSYS/Lumerical for the simulation software, Luceda Photonics for the IPKISS design software, and Thorlabs for various optomechanical hardware. We also thank the CORNERSTONE Project at the University of Southampton for fabrication of our chip designs.

\providecommand{\noopsort}[1]{}\providecommand{\singleletter}[1]{#1}%

\end{document}